\documentclass{iopjournal}
\usepackage{amsmath}
\usepackage{graphicx}
\usepackage{xcolor}
\usepackage[per-mode=fraction]{siunitx}
\usepackage{tabularx}
\usepackage{array}

\renewcommand{\articletype}[1]{{\vspace*{-8mm}\noindent \Large \sf Preprint}
	\vspace*{8mm} \\ \noindent
	{\scriptsize \sf{\bfseries \MakeUppercase{#1}}}}
\begin{document}

\articletype{Paper}

\title{Spectrotemporal imaging of ultrafast electron dynamics}

\author{D. Köhnke$^1$\orcid{0000-0003-1151-8625}, K. Ebken$^1$\orcid{0009-0006-6326-4005}, T. Bayer$^1$\orcid{0009-0005-2280-5672} and M. Wollenhaupt$^{1,*}$\orcid{0000-0002-0839-1494}}

\affil{$^1$Carl von Ossietzky Universität Oldenburg, Institute of Physics, Carl-von-Ossietzky-Straße 9-11, D-26129 Oldenburg}

\affil{$^*$Author to whom any correspondence should be addressed.}

\email{matthias.wollenhaupt@uol.de}


\begin{abstract}
We demonstrate a spectrotemporal imaging technique that maps ultrafast electron dynamics onto energy- and angle-resolved photoelectron distributions in a single measurement. The underlying principle is to encode the probing time into the kinetic energy of the emitted photoelectrons, while the evolving quantum dynamics is retrieved from their angular distributions. We experimentally demonstrate this concept by imaging spin-orbit wave packet dynamics in resonance-enhanced multiphoton ionization of potassium atoms using two complementary approaches. In the first, three spectrally distinct probe pulses sample the dynamics at selected delays and map the corresponding wave packet configurations onto separate photoelectron energy channels. In the second, a single broadband chirped probe pulse establishes a continuous time-to-energy mapping through its time-dependent instantaneous frequency, thereby encoding the evolution of the wave packet across a broad photoelectron energy band. The resulting photoelectron momentum distribution exhibits a characteristic fork-shaped pattern. Combining supercontinuum pulse shaping with velocity map imaging, we experimentally resolve the spin-orbit dynamics over a time window exceeding 300 fs, in good agreement with quantum dynamical simulations. Our results establish spectrotemporal photoelectron imaging as a versatile approach to time-resolved spectroscopy, replacing sequential delay scans by parallel acquisition of ultrafast quantum dynamics.		
\end{abstract}

\section{Introduction\label{Introduction}}

Ultrafast quantum dynamics are commonly investigated using pump-probe spectroscopy \cite{Zewail:2000:ACIE:2586,Mukamel:1995, Manz:1995,Dantus:2001:ARPC:639,Corkum:2007:NP:381,Kling:2008:ARPC:463,Wollenhaupt:2005:ARPC:25}, in which the time-evolution of a quantum system is reconstructed by systematically varying the time delay between pump and probe pulses. Although highly versatile, this approach is intrinsically sequential requiring separate measurements at different pump-probe delays, resulting in long acquisition times and making the experiment susceptible to drifts and shot-to-shot fluctuations. These limitations are particularly challenging in experiments with low repetition-rates, weak signals or limited sample stability.\\
Here, we demonstrate a conceptually different approach that maps the pump-probe delay onto the kinetic energy of the emitted photoelectrons, while encoding the ultrafast dynamics in the photoelectron momentum distribution (PMD).
Our approach builds on recent theoretical proposals for the spectrotemporal imaging of ultrafast dynamics using photoelectron spectroscopy \cite{Yudin:2006:PRL:063002, Jelovina:2017:PRA:043424}.
For its experimental implementation, we employ shaped femtosecond (fs) laser pulses which are well-established tools in coherent quantum control \cite{Rice:2000,Shapiro:2003,Vitanov:2001:ARPC:763,Tannor:2007,Silberberg:2009:ARPC:277,Weinacht:2021}. \\
The key idea is to combine laser pulse shaping for the generation of tailored probe pulses with photoelectron imaging spectroscopy for the differential detection of photoelectrons \cite{Weiner:2011:OC:3669,Wollenhaupt:2010:PRA:053422,Kerbstadt:2019:APX:1672583}. 
By tailoring the spectral and temporal properties of the probe field, different frequency components interact with the evolving quantum system at different times. 
Photoionization by different photon energies produces photoelectrons in different kinetic-energy windows, establishing a time-to-energy mapping. 
The quantum dynamics is independently encoded in the photoelectron angular distribution (PAD), allowing different stages of the evolution to be retrieved simultaneously from a single 3D PMD. In this way, the PMD provides a spectrotemporal multiplexed record of the dynamics without requiring a sequential delay scan.\\
Using two complementary implementations, we demonstrate the principle on ultrafast spin-orbit wave packet (SOWP) dynamics in the $4p$ fine-structure doublet of the potassium atom \cite{Zamith:2000:EPJD:255}. 
In the first, discrete implementation, we use spectral amplitude and linear spectral phase modulation of a white-light supercontinuum to generate a multichromatic pulse sequence consisting of a pump pulse followed by three probe pulses with distinct central wavelengths (colors) and delays. Each probe pulse interrogates the evolving SOWP at a different time and images the corresponding wave packet onto a separate photoelectron energy window. The coupled spin and orbital angular momentum dynamics manifest in the angular distribution of the released photoelectron partial wave packets \cite{Gilb:2006:JPB:4231,Wen:2006:PRA:052504,Goulielmakis:2010:Nature:739,Kerbstadt:2017:NJP:103017,Bayer:2019:NJP:033001, Li:2025:PRL:143201}. Using velocity map imaging (VMI) \cite{Eppink:1997:RSI:3477} and Abel inversion \cite{Garcia:2004:RSI:4989}, we reconstruct the PMD. In the discrete implementation, the reconstructed PMD contains three snapshots of the dynamics acquired simultaneously. In the second, continuous implementation, we replace the discrete probe pulse sequence with a single delayed, broadband, linearly chirped probe pulse. Its time-dependent instantaneous frequency establishes a continuous mapping between probing time and photon energy and, consequently, between probing time and photoelectron kinetic energy \cite{Yudin:2006:PRL:063002,Jelovina:2017:PRA:043424}. This implementation continuously images the SOWP evolution onto a broad photoelectron energy band, also capturing the temporal evolution between the discrete snapshots. \\
Building on the pioneering work of J. H. Eberly on coherent light-matter interactions and multilevel quantum dynamics, we support our experimental findings with numerical simulations \cite{Allen:1987,Eberly:1977:PRA:2038,Hioe:1984:PRA:1164,Shore:1978:OC:83,Younis:2022:JPBAMOP:164001}. Our simulations are based on the numerical solution of the time-dependent Schrödinger equation (TDSE) for an essential-state model of the potassium atom interacting with a shaped fs-pulse sequence. To account explicitly for electron spin and spin-orbit coupling, we describe the laser-atom interaction in a coupled angular-momentum basis.
\\
Our approach combines temporal multiplexing by shaped optical fields with spectroscopic decoding of ultrafast quantum dynamics, effectively replacing sequential delay scans with parallel acquisition of time-resolved information. By reducing sensitivity to slow experimental drifts and enabling efficient measurements at low repetition rates or weak signal levels, spectrotemporal imaging offers a versatile alternative to established pump-probe spectroscopy.
\section{Spectrotemporal Imaging Schemes}
\label{sec:PhysicalSystem}
\begin{figure}[htbp]
	\includegraphics[width=\linewidth]{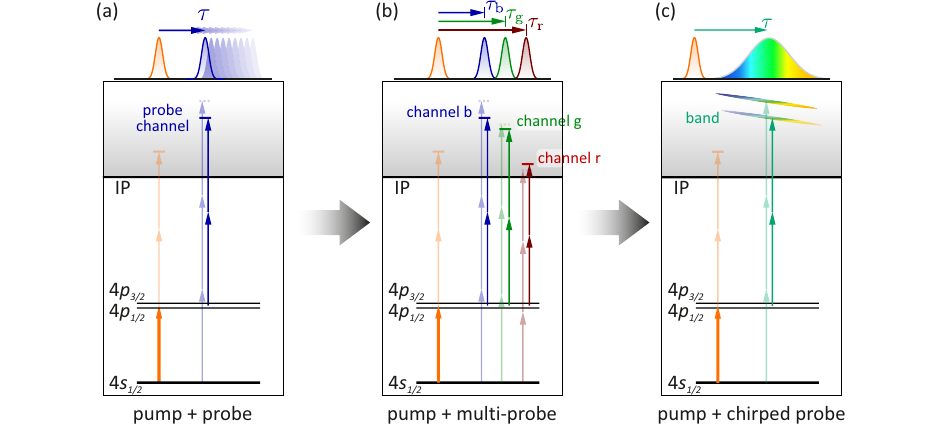}
	\caption{Excitation schemes for (a) the established pump-probe spectroscopy technique, (b) the discrete probing scheme and (c) the continuous probing scheme, for the imaging of ultrafast SOWP dynamics in the potassium $4p$ fine-structure doublet.
	While the traditional pump-probe technique maps the dynamics sequentially by varying the delay between the pump and probe pulse, the other two approaches enable spectrotemporal imaging of the dynamics in a single measurement using either a sequence of three probe pulses of different colors or a single chirped probe pulse. 
	The multipulse sequence images the SOWP dynamics at discrete times into different energy channels in the PMD.
	The chirped probe pulse images the dynamics continuously into a broad energy band in the PMD.
	\label{fig0:ExcScheme}}
\end{figure}
\begin{figure}[htbp]
	\includegraphics[width=\linewidth]{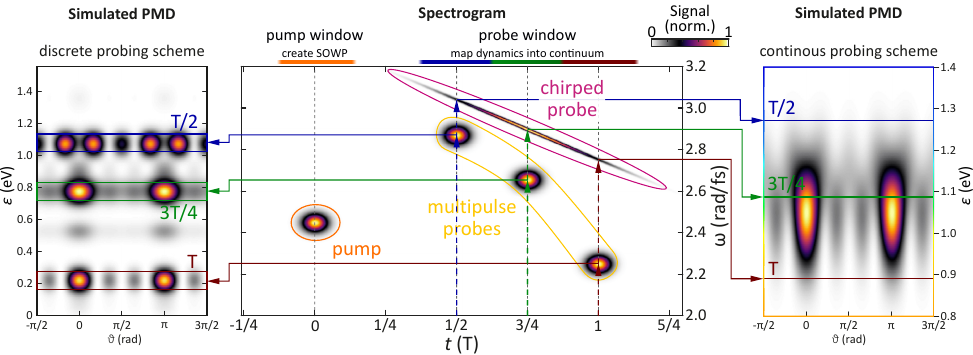}
	\caption{
	Schematic illustration of the discrete and continuous spectrotemporal imaging of the SOWP dynamics. The central panel shows spectrograms of the resonant pump pulse (orange), three discrete probe pulses (yellow), and a broadband chirped probe pulse (purple). The corresponding simulated PMDs are shown in the left and right panels. In the discrete probing scheme (left), three probe pulses with distinct central frequencies and delays map successive stages of the SOWP evolution onto separate photoelectron energy channels. In the continuous probing scheme (right), the frequency-dependent group delay of the chirped probe pulse maps the dynamics continuously onto a broad energy band, filling the temporal gaps between the discrete probe windows. The characteristic stages of the SOWP oscillation at $T/2$, $3T/4$ and $T$ are indicated for the corresponding photoelectron energy channels (left) and sections (right).}
	\label{fig1:ExcScheme}
\end{figure}

Figure~\ref{fig0:ExcScheme}(a) illustrates the established pump-probe scheme \cite{Weiner:2009:1} for the excitation of a SOWP in the potassium $4p$ fine-structure doublet by a resonant pump pulse. The ensuing electron dynamics is mapped onto the photoelectron spectrum via two-photon ionization by a variably delayed probe pulse. The nonlinear ionization step provides a narrower temporal gate because, in the perturbative regime, the $N$-photon ionization probability scales with the $N$th power of the intensity, resulting in an improved time resolution, roughly scaling with $\frac{1}{\sqrt{N}}$ for a Gaussian-type pulse. In addition, intermediate states generally play an important role in broadband multiphoton ionization \cite{Freeman:1987:PRL:1092,Krug:2009:NJP:105051,Bayer:2023:PRA:033111}, as recently investigated in detail by Eberly and co-workers \cite{Younis:2022:JPBAMOP:164001}. In the present study, the near-resonant potassium $4d$ states are of particular importance, as their population modifies the photoelectron signals and provide additional insight into the underlying ionization dynamics. We therefore pay special attention to the contributions of these intermediate states to the PMD, as discussed in Secs.~\ref{sec:ExperimentalSetup} and \ref{sec:Results}.\\
Our two extensions of the established pump-probe scheme towards discrete and continuous spectrotemporal mapping are illustrated in Figs.~\ref{fig0:ExcScheme}(b) and (c), respectively. Both implementations use the same pump pulse to launch the SOWP, and probe its time-evolution by two-photon ionization. Hence, all pump-probe schemes considered here are based on $(1+2)$ resonance-enhanced multiphoton ionization (REMPI), involving resonant excitation followed by two-photon ionization. In the discrete scheme shown in Fig.~\ref{fig0:ExcScheme}(b), three short probe pulses with different central frequencies (colors) map the dynamics at discrete delays $\tau_\mathrm{b}, \tau_\mathrm{g}, \tau_\mathrm{r}$ onto separate energy channels in the photoionization continuum. In the continuous scheme shown in Fig.~\ref{fig0:ExcScheme}(c), the discrete probe sequence is replaced by a single broadband, linearly chirped pulse. The resulting linear group delay establishes a continuous mapping between probing time and photoelectron kinetic energy, filling the temporal gaps between the three probe windows of the discrete scheme in (b). \\
The optical spectra of both schemes are designed to separate the relevant two-color pump-probe contributions (opaque arrows) from the single-color contributions (semi-transparent arrows) in the photoelectron spectrum, enabling the background-free observation of the dynamics. In addition, sufficient temporal separation of the subpulses is required to suppress frequency-mixing contributions that would otherwise congest the photoelectron spectrum.
The specific central wavelengths, spectral widths, and spectral phases employed in the experiment are detailed in Sec.~\ref{sec:ExperimentalSetup}.\\
Figure~\ref{fig1:ExcScheme} compares the discrete (left) and continuous (right) spectrotemporal probing schemes. Because the evolution of the bound SOWP manifests in the angular distribution of the emitted photoelectron wave packet \cite{Gilb:2006:JPB:4231,Wen:2006:PRA:052504,Bayer:2019:NJP:033001, Li:2025:PRL:143201}, we show two-dimensional sections through the simulated PMDs for both schemes. The PMDs were simulated by numerically solving the TDSE in the dipole approximation, using an essential-state model of potassium interacting with shaped fs-laser pulses \cite{Krug:2009:NJP:105051,Wollenhaupt:2009:APB:245,Bayer:2016:ACP:235}.
The model includes the seven bound states $\{4s_{1/2},4p_{1/2},4p_{3/2},3d_{3/2},3d_{5/2},4d_{3/2},4d_{5/2}\}$ and two discretized continua \cite{Charron:1998:JCP:3922,Rozgonyi:2000:JAP:2936,Lunden:2014:PRA:053403}, $\{\varepsilon f_{5/2},\varepsilon f_{7/2}\}$, each represented by 281 equidistant energy states. The eigenenergies of the bound states and the radial parts of the corresponding dipole couplings were taken from the NIST database \cite{Kramida:2018}. The couplings between the $d$-states and the $f$-type continua were treated perturbatively and assumed to be energy-independent (flat-continuum approximation). 
For the illustrative simulations in Fig.~\ref{fig1:ExcScheme}, the laser electric fields were generated in the frequency domain using Gaussian spectral amplitudes approximating the experimentally employed pulse spectra. For the subsequent quantitative comparison with the experimental photoelectron spectra, the measured optical spectra and the experimentally applied spectral phase functions were used directly in the simulations. The TDSE was solved numerically in the basis of the unperturbed atomic eigenstates using a short-time propagation method \cite{Tannor:2007,Bauer:2017}. To account for the finite energy resolution of the spectrometer, the simulated photoelectron spectra were convolved with a realistic Gaussian-shaped instrumental response function. Its full width at half maximum (FWHM) was set to \SI{80}{\milli\electronvolt} at a photoelectron kinetic energy of \SI{1}{\electronvolt}, corresponding to the measured energy resolution of our VMI spectrometer \cite{Pengel:2017:PRA:043426}. The temporal and spectral structure of the pulse sequences is visualized using a time-frequency representation. Several representations are available for characterizing ultrashort laser fields, including the Gabor spectrogram \cite{Gabor:1946:JIEEIII:429}, the Eberly-Wódkiewicz distribution \cite{Eberly:1977:JOSA:1252}, the Husimi distribution \cite{Husimi:1940:PPMSJ:264} and the Wigner distribution \cite{Wigner:1932:PR:749}; comprehensive overviews are given in \cite{Cohen:1995} and \cite{Auger:2008:131}. The central panel of Fig.~\ref{fig1:ExcScheme} shows the Wigner distributions of the individual subpulses
\begin{equation}
	\mathcal{W}_j(\omega,t) = \int E_j(t+\tau/2)E_j^*(t-\tau/2)\mathrm{e}^{\mathrm{i}\omega\tau}\mathrm{d}\tau,
\end{equation}
 where $E_j(t)$ denotes the field of the $j$th subpulse. For clarity, the Wigner distributions of the individual subpulses are calculated separately and superimposed, thereby excluding cross terms between different subpulses. The distributions $\mathcal{W}_j(\omega,t)$ are shown in a common time-frequency plane to facilitate direct comparison of the two probe schemes.
In both schemes, the pump pulse (orange ellipses) launches the $4p$-SOWP at $t=0$. The subsequent bound-state dynamics are therefore identical in both schemes and have been discussed in detail in \cite{Bayer:2019:NJP:033001}. The characteristic timescale is determined by the $4p$ fine-structure splitting of approximately $\Delta\varepsilon=\SI{7.15}{meV}$, corresponding to an oscillation period of $T=h/\Delta\varepsilon\approx\SI{578}{\femto\second}$. The discrete probing scheme employs three transform-limited probe pulses with distinct central wavelengths (encircled by the yellow line in Fig.~\ref{fig1:ExcScheme}). Each pulse probes the SOWP within a distinct time window and maps the corresponding stage of its evolution onto a separate photoelectron energy channel. The continuous probing scheme extends this concept by replacing the discrete probe pulses with a single broadband, down-chirped pulse. Its instantaneous frequency decreases with time, as illustrated by the corresponding Wigner distribution (purple ellipse), continuously mapping the SOWP evolution onto a broad photoelectron energy band. The chirped probe can thus be regarded as a continuous extension of the discrete probing scheme, capturing not only the snapshots sampled by the individual subpulses but also the intermediate stages of the SOWP evolution. In Fig.~\ref{fig1:ExcScheme}, the probe pulses of the discrete scheme are arranged in order of decreasing central frequency to highlight their correspondence with the down-chirped probe scheme. Unlike in the continuous scheme, however, their temporal ordering is not constrained by their central frequencies. Both the pulse ordering and the relative delays can be independently adjusted to suit the application. \\
In both simulated PMDs, the evolution of the SOWP manifests itself in changes in the PAD across the respective energy channels. In the discrete probing scheme, each channel exhibits a distinct PAD, evolving from a predominantly $|f,1\rangle$-type distribution in the upper channel (blue box) at $t=T/2$, through an intermediate distribution in the central channel (green box) at $t=3T/4$, to a pronounced $|f,0\rangle$-type distribution in the lower channel (red box) at $t=T$, after one complete period of the SOWP evolution. In the continuous probing scheme, the corresponding evolution appears as a smooth variation of the PAD across the photoelectron energy range covered by the probe signal. The mapping between probing time and photoelectron kinetic energy is governed by the instantaneous frequency of the chirped probe pulse,
\begin{equation}\label{eq:time_to_energy}
	\varepsilon(t) = \varepsilon_{4p} + 2\hbar\omega_\mathrm{inst}(t) - \varepsilon_\mathrm{IP},
\end{equation}
where $\varepsilon_{4p}$ denotes the mean energy of the $4p$ states, $\varepsilon_\mathrm{IP}$ is the ionization potential, and $\omega_\mathrm{inst}(t)$ is the instantaneous frequency of the chirped probe pulse. The factor of two accounts for the absorption of two probe photons in the ionization step \cite{Wollenhaupt:2010:PSUPSNMXIV:76000X}. Through this time-to-energy mapping, the PADs at the photoelectron energies associated with $t=T/2$, $3T/4$, and $T$ reproduce those obtained with the discrete probing scheme, as indicated by the blue, green, and red horizontal lines in the PMD.\\
\section{Experimental Setup and Strategy}
\label{sec:ExperimentalSetup}
In this section, we describe the experimental implementation and optical design of the two spectrotemporal imaging schemes introduced above. The experimental setup for white-light pulse shaping and photoelectron VMI has been described in detail elsewhere \cite{Pengel:2017:PRA:043426,Kerbstadt:2017:OE:12518}. Here, we focus on the components and pulse-shaping procedures relevant to the present study.\\
A white-light supercontinuum is generated by coupling \SI{25}{fs}, \SI{790}{\nano\meter} pulses from a chirped-pulse amplifier system (Femtolaser Femtopower HR, \SI{3}{\kilo\hertz}) into a neon-filled hollow-core fiber compressor at an absolute gas pressure of \SI{1.75}{\bar}.
The resulting broadband pulses are directed into a home-built $4f$ polarization pulse shaper (\textsc{Jenoptik SLM-S640d}) \cite{Kerbstadt:2017:JMO:1010, Kerbstadt:2017:OE:12518} for spectral phase and amplitude modulation.
Residual spectral phase compensation of the white-light pulses is achieved using established evolutionary optimization techniques to maximize the second-harmonic yield in a \SI{5}{\micro\meter}-thick $\beta$-barium borate (BBO) crystal \cite{Baumert:1997:APB:779,Yelin:1997:OL:1793,Kerbstadt:2017:JMO:1010}. The pump and probe spectral bands are then selected from the white-light supercontinuum by spectral amplitude modulation. Linear and quadratic spectral phase modulation functions are applied to the probe bands to control the relative delays of the probe pulses or the chirp rate of the probe pulse, respectively.\\
\begin{figure}[htbp]
	\centering
	\includegraphics[width=\linewidth]{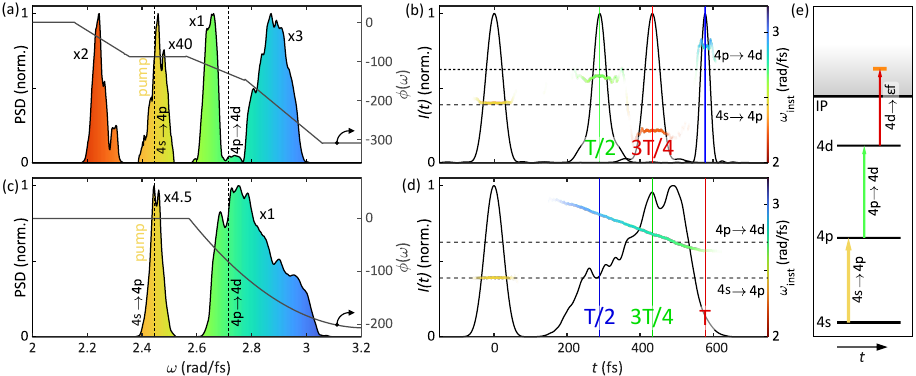}
	\caption{
	Spectral and temporal characteristics of the shaped fs-pulses employed in the the two spectrotemporal imaging schemes. (a,c) Measured PSDs and applied spectral phase functions (dark-gray lines) for the (a) discrete multipulse and (c) continuous chirped-pulse scheme, respectively. (b,d) Corresponding calculated temporal intensity profiles (black lines), with the instantaneous optical frequencies of the individual pulses indicated by colored lines. The horizontal dashed lines indicate the frequencies of the $4s\rightarrow4p$ and $4p\rightarrow4d$ transitions. (e) Simplified excitation scheme illustrating the pump excitation, resonant intermediate $4d$ state, and subsequent probe ionization mapping the dynamics into the continuum.
	}
	\label{fig2:OpticalPreparation}
\end{figure}
Figure~\ref{fig2:OpticalPreparation} shows the measured power spectral densities (PSDs) and applied spectral phase functions (left panels), together with the calculated temporal intensity profiles and corresponding instantaneous frequencies (center panels). Panels (a,b) illustrate the discrete probing scheme, while panels (c,d) show the continuous probing scheme. For clarity, the individual spectral bands are normalized separately, with the corresponding normalization factors indicated in the figure; the temporal pulses are likewise normalized.\\
In the discrete probing scheme, four spectral bands were extracted from the white-light supercontinuum. Their central wavelengths and bandwidths are chosen to ensure that the individual (1+2) REMPI contributions are energetically separated in the PMD, both from one another and from the corresponding single-color three-photon ionization signals. At the same time, the spectral bandwidths are sufficiently broad compared to the $4p$ fine-structure splitting to enable impulsive excitation and probing of the SOWP dynamics. The yellow spectral band, which forms the pump pulse, is centered at the $4s\rightarrow4p$ transition frequency of $\omega_\mathrm{pump}\approx\SI[per-mode=symbol]{2.45}{\radian\per\femto\second}$. The pump bandwidth corresponds to a transform-limited pulse duration of approximately \SI{60}{\femto\second}. The pump amplitude is kept sufficiently low to suppress single-color photoelectron emission via (1+2) REMPI from the $4s$ ground state. Each probe pulse has a transform-limited duration of approximately \SI{35}{\femto\second}, providing a short temporal window for probing the SOWP dynamics. The spectral components of the green probe band near \SI[per-mode=symbol]{2.71}{\radian\per\femto\second}, resonant with the $4p\rightarrow4d$ transition, were deliberately retained to illustrate the influence of intermediate resonances, which generally have to be taken into account in multiphoton probing schemes \cite{Younis:2022:JPBAMOP:164001}. The green probe pulse therefore resonantly drives the $4p\rightarrow4d$ transition, transferring a fraction of the $4p$ population to the $4d_{3/2}$ and $4d_{5/2}$ states. To demonstrate the flexibility of the discrete probing scheme discussed in Sec.~\ref{sec:PhysicalSystem}, we rearranged the probe pulses into a green-red-blue sequence, rather than the descending-frequency sequence shown in Fig.~\ref{fig1:ExcScheme}. In this sequence, the green pulse precedes the red pulse, allowing population transferred to the $4d$ states by the green pulse to be subsequently mapped onto a separate photoelectron energy band by the red pulse. The corresponding (1+1+1) REMPI pathway is illustrated in the excitation scheme in Fig.~\ref{fig2:OpticalPreparation}(e), where fine-structure splittings are omitted for clarity. A second consideration in choosing the pulse ordering is the non-resonant three-photon ionization induced by the green pulse. This single-color contribution lies close in photoelectron energy to the corresponding (1+2) REMPI signal. To distinguish these contributions by their angular distributions, the green probe pulse is timed at $\tau_\mathrm{g}=T/2$, where the PAD of the two-color contribution originating from the $4p$ states is expected to differ most strongly from that of the adjacent single-color contribution originating directly from the $4s$ ground state. The spectral amplitudes of the probe pulses were adjusted to obtain comparable photoelectron yields across the different energy channels.\\
In the continuous probing scheme, the white-light supercontinuum is divided into a pump band and a single broadband probe band. The pump spectral band was similar to that used in the discrete probing scheme, while the probe band was centered around \SI[per-mode=symbol]{2.81}{\radian\per\femto\second}. Its spectral bandwidth was chosen to maximize the photoelectron energy range accessible to spectrotemporal imaging, limited by the remaining bandwidth of the white-light supercontinuum. The corresponding transform-limited duration of the probe pulse was approximately \SI{15}{\femto\second}. A linear spectral phase $\varphi(\omega)=\phi^{(1)}(\omega-\omega_\mathrm{mod})$ was applied to delay the probe pulse by $\phi^{(1)}=\SI{370}{\femto\second}$ relative to the pump. An additional quadratic spectral phase $\varphi(\omega)=\frac{1}{2}\phi^{(2)}(\omega-\omega_\mathrm{mod})^2$, with $\phi^{(2)}=\SI{-1000}{\femto\second\squared}$, introduced a negative chirp and stretched the probe pulse to approximately \SI{315}{\femto\second} (full width at $1/e$ of the peak intensity), exceeding half the SOWP period ($T/2$). To suppress the single-color photoelectron contribution from non-resonant three-photon ionization of the $4s$ ground state by the probe pulse, the pump amplitude was chosen sufficiently high to achieve substantial ground-state depletion ($\pi$-pulse excitation). As a result, the two-color pump-probe signal exceeded the contribution from direct ground-state ionization by the probe pulse. The phase parameters $\phi^{(1)}$ and $\phi^{(2)}$ were chosen such that the probe pulse covered approximately the second half of the SOWP period while remaining sufficiently separated from the pump pulse to suppress additional PMD contributions arising from non-resonant two-color frequency mixing. Since the SOWP dynamics were probed through two-photon ionization, intermediate resonances introduced an asymmetry in the (1+2) REMPI process with respect to the sign of the chirp. Depending on the sign of the chirp, the $4p\rightarrow4d$ transition was driven resonantly at different times during the probe pulse. For a down-chirped probe, the resonance was reached toward the end of the pulse, as indicated by the intersection of the instantaneous frequency with the horizontal dashed line marking the transition frequency in Fig.~\ref{fig2:OpticalPreparation}(d). For an up-chirped probe (not shown), in contrast, the resonance would have been reached near the beginning of the pulse. In this case, population transferred resonantly to the $4d$ states could have been further ionized by the remaining part of the probe pulse, producing an unwanted (1+1+1) REMPI contribution that would overlap with the desired (1+2) REMPI signal from the $4p$ states, analogous to the pathway illustrated in Fig.~\ref{fig2:OpticalPreparation}(e). Therefore, a down-chirped probe pulse was employed in the experiment to minimize the influence of the intermediate $4p\rightarrow4d$ resonance and enable a clear demonstration of the continuous probing scheme.\\
\section{Results and Discussion}
\label{sec:Results}
\begin{figure}
	\includegraphics[width=\linewidth]{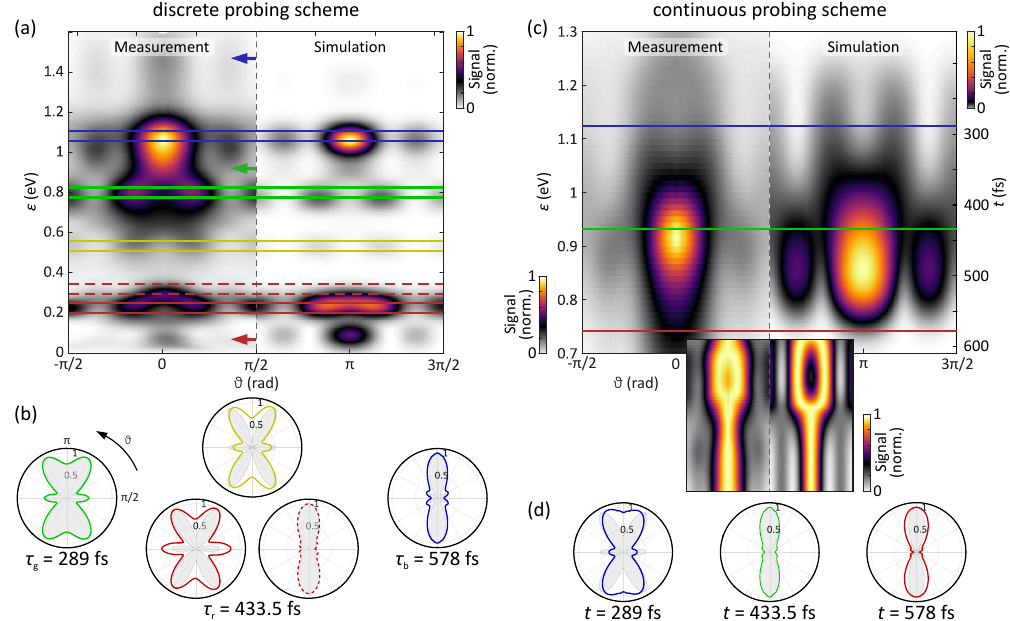}
	\caption{
	Experimental and simulated spectrotemporal imaging of SOWP dynamics using the discrete and the continuous probing schemes. (a) Abel-inverted experimental (left) and simulated (right) PMDs for the discrete scheme. The colored lines indicate the different photoelectron energy channels. (b) Corresponding normalized PADs extracted by integration over the selected energy channels. Colored lines show the experimental data and gray-shaded curves represent the simulations. (c) Experimental (left) and simulated (right) PMDs for the continuous probing scheme. The right-hand axis indicates the probing time obtained from the time-to-energy mapping. The inset shows the experimental PADs normalized independently at each photoelectron energy, revealing the characteristic fork-shaped pattern of the evolving SOWP. (d) PADs extracted at selected energies corresponding to probing times of $T/2$, $3T/4$, and $T$.
	\label{fig3:MainResults}
	}
\end{figure}

The experimental results are presented in Fig.~\ref{fig3:MainResults} alongside the corresponding numerical simulations. The only adjustable parameters in the simulations were the relative amplitudes of the subpulses, which were adjusted to reproduce the measured PMDs.
The upper row shows the measured and simulated PMDs for (a) the discrete multipulse scheme and (c) the continuous chirped-pulse scheme. The lower row presents the corresponding normalized PADs, obtained by integrating over the individual photoelectron energy channels for the discrete scheme in (b) and by extracting angular distributions at selected photoelectron energies for the continuous scheme in (d). In both schemes, the selected photoelectron energies correspond to specific probing times and thus to distinct stages of the SOWP dynamics.\\
Figure~\ref{fig3:MainResults}(a) shows the measured, symmetrized, and Abel-inverted PMD for the discrete probing scheme (left) alongside the corresponding simulation (right). Both exhibit several well-separated photoelectron energy channels with distinct angular distributions. The pump-probe contributions from two-color $(1+2)$ REMPI are marked by the green ($\tau_\mathrm{g}=\SI{289}{\femto\second}=T/2$), red ($\tau_\mathrm{r}=\SI{433.5}{\femto\second}=3T/4$), and blue ($\tau_\mathrm{b}=\SI{578}{\femto\second}=T$) lines, indicating the corresponding probing times. The PADs extracted from these channels, shown in Fig.~\ref{fig3:MainResults}(b), allow a direct comparison of the measured dynamics with the theoretically predicted time-evolution discussed in detail in \cite{Bayer:2019:NJP:033001}.
The experimental PADs agree closely with the simulated distributions, shown as gray-shaded backgrounds in Fig.~\ref{fig3:MainResults}(b).
In particular, the green channel at $\tau_\mathrm{g}=T/2$ exhibits the expected $|f,1\rangle$-type distribution, whereas the blue channel at $\tau_\mathrm{b}= T$ shows the characteristic $|f,0\rangle$-type distribution.
The PAD observed in the intermediate time-window around $\tau_\mathrm{r}=3T/4$ captures the transition between these two limiting cases. \\
In addition to these pump-probe contributions, several other features are observed in the PMD. Signals at approximately \SI{0.1}{\electronvolt}, \SI{0.9}{\electronvolt}, and \SI{1.5}{\electronvolt}, indicated in Fig.~\ref{fig3:MainResults}(a) by the red, green and blue arrow, respectively, are attributed to non-resonant three-photon single-color ionization of the $4s$ ground state by the corresponding probe pulses. The red pump-probe channel is split into two contributions in the measured PMD, marked by the solid and dashed red lines. The lower-energy component around \SI{0.2}{\electronvolt} shows a predominant $|f,1\rangle$-type angular distribution, whereas the higher-energy component around \SI{0.3}{\electronvolt} resembles an $|f,0\rangle$-type distribution.The observed splitting is reproduced in the simulations when the amplitude of the red probe pulse is increased. Analysis of the simulated population dynamics indicates that the two contributions arise from non-resonant population oscillations in the $4s$-$4p$ subsystem driven by the comparatively intense red probe pulse. 
An additional contribution appears near \SI{0.5}{\electronvolt}, marked by the yellow lines, which originates from the three-color (1+1+1) REMPI pathway illustrated in Fig.~\ref{fig2:OpticalPreparation}(e). 
The PAD in this yellow channel resembles that of the green channel, despite the different probing times of $3T/4$ and $T/2$, respectively. 
The similarity arises from the resonance-enhanced two-photon ionization of the $4p$ states via the $4d$ states by the green pulse.
During the ionization, some population is stored in the $4d$ doublet, effectively conserving the SOWP configuration of the $4p$ states at $T/2$.
Because the $4d$ fine-structure splitting is only \SI{0.133}{\milli\electronvolt}, corresponding to a SOWP period of approximately \SI{31}{\pico\second}, the subsequent evolution is negligible during the time between the green and red probe pulses.
Consequently, the red pulse arriving at $3T/4$ images an essentially unchanged SOWP configuration from the $4d$ states into the yellow channel, closely resembling the configuration at $T/2$.
This explains the similarity of the resulting PADs in the green and yellow channels produced by ionization via the $4d$ states by the green and red pulse, respectively.
\\
Figure~\ref{fig3:MainResults}(c) shows the measured and simulated PMDs for the continuous probing scheme. For the down-chirped probe used here, the higher-frequency components arrive earlier and thus map earlier stages of the SOWP dynamics onto higher photoelectron kinetic energies, whereas later stages appear at lower energies. 
The time axis shown on the right-hand side of Fig.~\ref{fig3:MainResults}(c) is obtained by inverting the time-to-energy relation in Eq.~\eqref{eq:time_to_energy}.
The instantaneous frequency entering this relation was determined from a linear fit to the curve shown in Fig.~\ref{fig2:OpticalPreparation}(d). The resulting time-to-energy mapping is therefore specific to the spectral phase and shape of the probe pulse. The experimental PMD displays the expected continuous evolution of the PAD across the probe energy band, extending approximately from \SI{0.7}{eV} to \SI{1.3}{eV}. 
With the phase parameters ($\phi^{(1)},\phi^{(2)}$) chosen to map the second half of the SOWP period onto the photoelectron energy axis, the PAD evolves from an $|f,1\rangle$-type distribution at $\varepsilon=\SI{1.12}{eV}$, corresponding to $t=T/2$ (blue solid line), towards an $|f,0\rangle$-type distribution at $\varepsilon=\SI{0.74}{eV}$, corresponding to $t=T$ (red solid line). The continuous evolution between these characteristic angular distributions gives rise to a distinctive fork-shaped pattern in the PMD, most clearly visible in the inset of Fig.~\ref{fig3:MainResults}(c). Here, the PADs are normalized independently at each photoelectron energy, suppressing variations in the total photoelectron yield arising from the spectral intensity distribution of the chirped probe pulse. This normalization highlights the angular signature of the evolving SOWP. The photoelectron signal reaches its maximum near $\varepsilon=\SI{0.93}{\electronvolt}$, corresponding to $t=3T/4$. The PADs extracted at these energies, shown in Fig.~\ref{fig3:MainResults}(d), agree with those obtained at the corresponding probing times in the discrete probing scheme. Beyond these selected delays, the chirped probe maps the intermediate stages of the SOWP evolution continuously onto the photoelectron energies in between. 
The simulation reproduces the characteristic evolution of the PMD observed in the experiment, with the color scale slightly adjusted to highlight the agreement between the angular distributions.
Small discrepancies remain in the PADs, with less pronounced nodes in the measured distributions, most notably at $t=T/2$. These discrepancies may arise from the omission of ionization into the $p$-type continuum \cite{Koehnke:2023:NJP:123025} in the simulations and, in part, from an additional experimental background.\\
Finally, we compare the temporal resolution of the two probing schemes. In the discrete scheme, the temporal resolution is primarily determined by the effective duration of the individual probe pulses, as in conventional pump-probe spectroscopy. In the continuous scheme, temporal resolution is linked to energy resolution through the time-to-energy mapping $\varepsilon(t)$ (see Eq.~\eqref{eq:time_to_energy}).
Mapping the \SI{80}{\milli\electronvolt} energy resolution of our VMI spectrometer (cf. Sec.~\ref{sec:PhysicalSystem}) back onto the time axis yields a temporal resolution of approximately \SI{60}{\femto\second} for the chirp parameter used here. Thus, the temporal resolution of the two schemes is governed by different experimental parameters: the probe pulse duration in the discrete scheme and the combination of spectrometer energy resolution and time-to-energy mapping in the continuous scheme.\\
The results presented here demonstrate two shaper-based approaches to scan-free spectrotemporal imaging of ultrafast electron dynamics. The discrete probing scheme allows the timing, central frequencies, and bandwidths of the individual probe pulses to be controlled independently, enabling intermediate resonances to be selectively addressed or avoided. In contrast, the continuous probing scheme maps an extended time interval onto the photoelectron energy axis using a single chirped probe pulse, providing access to the evolving dynamics without requiring additional probe pulses. Both approaches encode the temporal evolution of the electron dynamics in the photoelectron momentum distribution, allowing different stages to be retrieved from a single measurement.\\
\section{Conclusion}
\label{sec:Conclusion}
We have presented an experimental technique for the direct scan-free spectrotemporal imaging of ultrafast dynamics in a single pump-probe measurement.
The key idea is to utilize programmable ultrafast pulse shaping for the generation of tailored probe pulses that map a range of pump-probe time-delays onto the kinetic energy distribution of the released photoelectrons.
The dynamical information is encoded in the PAD where it can be retrieved using differential photoelectron detection. \\
We demonstrated the concept on ultrafast SOWP dynamics in (1+2) REMPI of potassium atoms. Employing supercontinuum pulse shaping, we implemented two different probing schemes. In the first scheme, a short resonant pump pulse was followed by three probe pulses of different color and delay, imaging the SOWP dynamics discretely into three separate energy channels in the PMD. In the second scheme, a single broadband chirped probe pulse was used to sample the SOWP time evolution continuously and image the dynamics into a broad energy window in the PMD. Specifically, we mapped a time-window of more than \SI{300}{fs} onto the radial part of the PMD. The dynamics encoded in the angular part were measured using VMI techniques and reconstructed by Abel inversion. The continuous evolution of the PADs gives rise to a characteristic fork-shaped pattern in the energy-resolved PMD, providing a direct visual signature of the mapped SOWP dynamics. Our experimental findings are supported by numerical TDSE simulations that explicitly account for the electron spin and spin-orbit coupling.\\
The temporal resolution of the two schemes is determined by different experimental parameters. In the discrete probing scheme, the attainable resolution is mainly limited by the probe pulse duration, whereas in the continuous scheme, it depends on the energy resolution of the VMI spectrometer through the time-to-energy mapping.
In both cases, the nonlinear probing process can improve the temporal resolution. However, our work also shows that the influence of intermediate states, inevitably encountered in broadband multiphoton ionization, requires careful consideration. Near-resonant intermediate states can give rise to additional photoelectron contributions and modify the main contributions, potentially complicating their interpretation. \\
More generally, our results illustrate the potential of programmable pulse shaping beyond the generation of precisely tailored multicolor pulse sequences. The ability to design probe fields with specific temporal and spectral properties, such as chirped pulses, provides additional control over the light-matter interaction and the underlying excitation and ionization pathways. This flexibility enables the investigation of dynamical mechanisms and opens up experimental techniques that are not readily accessible with conventional sequences of short pulses. \\
Our shaper-based technique for spectrotemporal imaging of ultrafast dynamics is particularly suited to experiments with low repetition rates, weak signals, or limited sample stability, providing a route towards scan-free time-resolved photoelectron spectroscopy.
In particular, the short-pump-chirped-probe scheme is attractive for applications in XUV photoelectron spectroscopy, e.g., using synchrotron radiation sources \cite{Kaneyasu:2026:PRA:043104}, free-electron lasers \cite{Prince:2016:NP:176}, or attosecond laser pulses from high-harmonic radiation \cite{Cattaneo:2016:OE:29060}. At free-electron laser facilities such as FERMI, the combination of a short pump pulse with a deliberately chirped XUV probe pulse could enable spectrotemporal imaging of ultrafast dynamics without a sequential delay scan. Such approaches are particularly promising where stable optomechanical delay lines are difficult to implement, while techniques for generating chirped pulses are emerging \cite{Yudin:2006:PRL:063002,Richter:2024:Nature:337}.

\funding{This work was funded by the collaborative research program \textit{Dynamics on the Nanoscale} (DyNano) and the Wissenschaftsraum \textit{Elektronen-Licht-Kontrolle} (elLiKo) funded by the Niedersächsische Ministerium für Wissenschaft und Kultur.}

\data{The datasets used during the current study are available from the corresponding author on reasonable request.}

\bibliographystyle{ieeetr}
\bibliography{ultra_db}

\begin{thebibliography}{10}

\bibitem{Zewail:2000:ACIE:2586}
A.~H. Zewail, ``Femtochemistry: Atomic-scale dynamics of the chemical bond
  using ultrafast lasers (nobel lecture),'' {\em Angew. Chem. Int. Ed.},
  vol.~39, no.~15, pp.~2586--2631, 2000.

\bibitem{Mukamel:1995}
S.~Mukamel, {\em Principles of nonlinear optical spectroscopy}, vol.~6.
\newblock Oxford university press New York, 1995.

\bibitem{Manz:1995}
J.~Manz and L.~Woeste, {\em Femtosecond Chemistry}.
\newblock Weinheim: VCH, 1995.

\bibitem{Dantus:2001:ARPC:639}
M.~Dantus, ``Coherent nonlinear spectroscopy: From femtosecond dynamics to
  control,'' {\em Annu. Rev. Phys. Chem.}, vol.~52, pp.~639--589, 2001.

\bibitem{Corkum:2007:NP:381}
P.~B. Corkum and F.~Krausz, ``Attosecond science,'' {\em Nat. Phys.}, vol.~3,
  pp.~381--387, 2007.

\bibitem{Kling:2008:ARPC:463}
M.~F. Kling and M.~J.~J. Vrakking, ``Attosecond electron dynamics,'' {\em Annu.
  Rev. Phys. Chem.}, vol.~59, pp.~463--499, 2008.

\bibitem{Wollenhaupt:2005:ARPC:25}
M.~Wollenhaupt, V.~Engel, and T.~Baumert, ``Femtosecond laser photoelectron
  spectroscopy on atoms and small molecules: Prototype studies in quantum
  control,'' {\em Annu. Rev. Phys. Chem.}, vol.~56, pp.~25--56, 2005.

\bibitem{Yudin:2006:PRL:063002}
G.~L. Yudin, A.~D. Bandrauk, and P.~B. Corkum, ``Chirped attosecond
  photoelectron spectroscopy,'' {\em Phys. Rev. Lett.}, vol.~96, no.~6,
  p.~063002, 2006.

\bibitem{Jelovina:2017:PRA:043424}
D.~Jelovina, J.~Feist, F.~Mart\'{i}n, and A.~Palacios, ``Imaging ultrafast
  molecular wave packets with a single chirped uv pulse,'' {\em Phys. Rev. A},
  vol.~95, no.~4, p.~043424, 2017.

\bibitem{Rice:2000}
S.~A. Rice and M.~Zhao, {\em Optical Control of Molecular Dynamics}.
\newblock New York: Wiley, 2000.

\bibitem{Shapiro:2003}
M.~Shapiro and P.~Brumer, {\em Principles of the Quantum Control of Molecular
  Processes}, vol.~1.
\newblock Hoboken, New Jersey: John Wiley and Sons, 2003.

\bibitem{Vitanov:2001:ARPC:763}
N.~V. Vitanov, T.~Halfmann, B.~W. Shore, and K.~Bergmann, ``Laser-induced
  population transfer by adiabatic passage techniques,'' {\em Annu. Rev. Phys.
  Chem.}, vol.~52, pp.~763--809, 2001.

\bibitem{Tannor:2007}
D.~Tannor, {\em Introduction to Quantum Mechanics: A Time-Dependent
  Perspective}.
\newblock Sausalito: University Science Books, 2007.

\bibitem{Silberberg:2009:ARPC:277}
Y.~Silberberg, ``Quantum coherent control for nonlinear spectroscopy and
  microscopy,'' {\em Annu. Rev. Phys. Chem.}, vol.~60, pp.~277--292, 2009.

\bibitem{Weinacht:2021}
T.~C. Weinacht and B.~J. Pearson, {\em Time-Resolved Spectroscopy: An
  Experimental Perspective}.
\newblock Textbook Series in Physical Sciences, Boca Raton, FL 33487-2742: CRC
  Press, 2021.

\bibitem{Weiner:2011:OC:3669}
A.~M. Weiner, ``Ultrafast optical pulse shaping: A tutorial review,'' {\em Opt.
  Comm.}, vol.~284, pp.~3669--3692, 2011.

\bibitem{Wollenhaupt:2010:PRA:053422}
M.~Wollenhaupt, T.~Bayer, N.~V. Vitanov, and T.~Baumert, ``Three-state
  selective population of dressed states via generalized spectral phase-step
  modulation,'' {\em Phys. Rev. A}, vol.~81, p.~053422, 2010.

\bibitem{Kerbstadt:2019:APX:1672583}
S.~Kerbstadt, K.~Eickhoff, T.~Bayer, and M.~Wollenhaupt, ``Control of free
  electron wave packets by polarization-tailored ultrashort bichromatic laser
  fields,'' {\em Adv. Phys.: X}, vol.~4, no.~1, p.~1672583, 2019.

\bibitem{Zamith:2000:EPJD:255}
S.~Zamith, M.~A. Bouchene, E.~Sokell, C.~Nicole, V.~Blanchet, and B.~Girard,
  ``Observation of the oscillation of an angular wavepacket,'' {\em Eur. Phys.
  J. D}, vol.~12, pp.~255--261, 2000.

\bibitem{Gilb:2006:JPB:4231}
S.~Gilb, E.~A. Torres, and S.~R. Leone, ``Mapping of time-dependent electron
  orbital alignment,'' {\em J. Phys. B}, vol.~39, no.~20, p.~4231, 2006.

\bibitem{Wen:2006:PRA:052504}
H.~Wen, S.~N. Pisharody, J.~M. Murray, and P.~H. Bucksbaum, ``Observing angular
  precession of a rydberg wave packet due to spin-orbit coupling by
  orthogonally polarized weak half-cycle pulses,'' {\em Phys. Rev. A}, vol.~73,
  p.~052504, 2006.

\bibitem{Goulielmakis:2010:Nature:739}
E.~Goulielmakis, Z.~H. Loh, A.~Wirth, R.~Santra, N.~Rohringer, V.~S. Yakovlev,
  S.~Zherebtsov, T.~Pfeifer, A.~M. Azzeer, M.~F. Kling, S.~R. Leone, and
  F.~Krausz, ``Real-time observation of valence electron motion,'' {\em
  Nature}, vol.~466, pp.~739--743, 2010.

\bibitem{Kerbstadt:2017:NJP:103017}
S.~Kerbstadt, D.~Pengel, D.~Johannmeyer, L.~Englert, T.~Bayer, and
  M.~Wollenhaupt, ``Control of photoelectron momentum distributions by
  bichromatic polarization-shaped laser fields,'' {\em New J. Phys.}, vol.~19,
  p.~103017, 2017.

\bibitem{Bayer:2019:NJP:033001}
T.~Bayer, D.~Gr\"{a}fing, S.~Kerbstadt, D.~Pengel, K.~Eickhoff, L.~Englert, and
  M.~Wollenhaupt, ``Time-resolved {3D} imaging of ultrafast spin-orbit wave
  packet dynamics,'' {\em New J. Phys.}, vol.~21, no.~3, p.~033001, 2019.

\bibitem{Li:2025:PRL:143201}
W.~Li, J.~Zhang, Y.~Jin, D.~Zhang, D.~Ding, and K.~Ueda, ``Partial-wave
  resolved spin-orbit dynamics,'' {\em Phys. Rev. Lett.}, vol.~135, no.~14,
  p.~143201, 2025.

\bibitem{Eppink:1997:RSI:3477}
A.~T. J.~B. Eppink and D.~H. Parker, ``Velocity map imaging of ions and
  electrons using electrostatic lenses: Application in photoelectron and
  photofragment ion imaging of molecular oxygen,'' {\em Rev. Sci. Instrum.},
  vol.~68, no.~9, pp.~3477--3484, 1997.

\bibitem{Garcia:2004:RSI:4989}
G.~A. Garcia, L.~Nahon, and I.~Powis, ``Two-dimensional charged particle image
  inversion using a polar basis function expansion,'' {\em Rev. Sci. Instrum.},
  vol.~75, no.~11, pp.~4989--4996, 2004.

\bibitem{Allen:1987}
L.~Allen and J.~H. Eberly, {\em Optical Resonance and Two-Level Atoms}, vol.~2.
\newblock New York: Dover Publications, 1987.

\bibitem{Eberly:1977:PRA:2038}
J.~H. Eberly, B.~W. Shore, Z.~Białynicka-Birula, and I.~Białynicki-Birula,
  ``Coherent dynamics of $n$-level atoms and molecules. i. numerical
  experiments,'' {\em Phys. Rev. A}, vol.~16, no.~5, pp.~2038--2047, 1977.

\bibitem{Hioe:1984:PRA:1164}
F.~T. Hioe and J.~H. Eberly, ``Multiple-laser excitation of multilevel atoms,''
  {\em Phys. Rev. A}, vol.~29, no.~3, pp.~1164--1167, 1984.

\bibitem{Shore:1978:OC:83}
B.~W. Shore and J.~H. Eberly, ``Analytic approximations in multi-level
  excitation theory,'' {\em Optics Communications}, vol.~24, no.~1, pp.~83--88,
  1978.

\bibitem{Younis:2022:JPBAMOP:164001}
D.~Younis and J.~H. Eberly, ``Benchmark of few-level quantum theory vs ab
  initio numerical solutions for the strong-field autler-townes effect in
  photoionization of hydrogen,'' {\em J. Phys. B: At. Mol. Opt. Phys.},
  vol.~55, no.~16, p.~164001, 2022.

\bibitem{Weiner:2009:1}
A.~M. Weiner, {\em Ultrafast Optics}.
\newblock Hoboken, New Jersey: John Wiley \& Sons, 2009.

\bibitem{Freeman:1987:PRL:1092}
R.~R. Freeman, P.~H. Bucksbaum, H.~M. Milchberg, S.~Darack, D.~Schumacher, and
  M.~E. Geusic, ``Above-threshold ionization with subpicosecond laser pulses,''
  {\em Phys. Rev. Lett.}, vol.~59, no.~10, p.~1092, 1987.

\bibitem{Krug:2009:NJP:105051}
M.~Krug, T.~Bayer, M.~Wollenhaupt, C.~Sarpe-Tudoran, T.~Baumert, S.~S. Ivanov,
  and N.~V. Vitanov, ``Coherent strong-field control of multiple states by a
  single chirped femtosecond laser pulse,'' {\em New J. Phys.}, vol.~11,
  p.~105051, 2009.

\bibitem{Bayer:2023:PRA:033111}
T.~Bayer, K.~Eickhoff, D.~K\"{o}hnke, and M.~Wollenhaupt, ``Phase control of
  the autler-townes doublet in multistate systems,'' {\em Phys. Rev. A},
  vol.~108, no.~3, p.~033111, 2023.

\bibitem{Wollenhaupt:2009:APB:245}
M.~Wollenhaupt, M.~Krug, J.~K\"{o}hler, T.~Bayer, C.~Sarpe-Tudoran, and
  T.~Baumert, ``Photoelectron angular distributions from strong-field coherent
  electronic excitation,'' {\em Appl. Phys. B}, vol.~95, pp.~245--259, 2009.

\bibitem{Bayer:2016:ACP:235}
T.~Bayer, M.~Wollenhaupt, H.~Braun, and T.~Baumert, ``Ultrafast and efficient
  control of coherent electron dynamics via {SPODS},'' {\em Adv. Chem. Phys.},
  vol.~159, p.~235, 2016.

\bibitem{Charron:1998:JCP:3922}
E.~Charron and A.~Suzor-Weiner, ``Femtosecond dynamics of nai ionization and
  dissociative ionization,'' {\em J. Chem. Phys.}, vol.~108, no.~10,
  pp.~3922--3931, 1998.

\bibitem{Rozgonyi:2000:JAP:2936}
T.~Rozgonyi, A.~Gla{\ss}, and T.~Feurer, ``Application of nonreflecting
  boundary condition for numerical simulation of molecular photoionization
  dynamics,'' {\em J. Appl. Phys.}, vol.~88, no.~5, pp.~2936--2942, 2000.

\bibitem{Lunden:2014:PRA:053403}
W.~D.~M. Lunden, P.~S\'{a}ndor, T.~C. Weinacht, and T.~Rozgonyi, ``Model for
  describing resonance-enhanced strong-field ionization with shaped ultrafast
  laser pulses,'' {\em Phys. Rev. A}, vol.~89, p.~053403, 2014.

\bibitem{Kramida:2018}
A.~Kramida, Y.~Rachenko, and J.~Reader, ``{NIST} atomic spectra database,''
  {\em National Institute of Standards and Technology}, vol.~The online
  database is available at https://www.nist.gov/pml/atomic-spectra-database,
  2018.

\bibitem{Bauer:2017}
D.~Bauer, {\em Computational strong-field quantum dynamics. Intense
  light-matter interactions}, vol.~1.
\newblock Berlin: De Gruyter, 2017.

\bibitem{Pengel:2017:PRA:043426}
D.~Pengel, S.~Kerbstadt, L.~Englert, T.~Bayer, and M.~Wollenhaupt, ``Control of
  three-dimensional electron vortices from femtosecond multiphoton
  ionization,'' {\em Phys. Rev. A}, vol.~96, p.~043426, 2017.

\bibitem{Gabor:1946:JIEEIII:429}
D.~Gabor, ``Theory of communication. part 1: The analysis of information,''
  {\em Journal of the Institution of Electrical Engineers-part III: radio and
  communication engineering}, vol.~93, no.~26, pp.~429--441, 1946.

\bibitem{Eberly:1977:JOSA:1252}
J.~H. Eberly and K.~W\'{o}dkiewicz, ``The time-dependent physical spectrum of
  light,'' {\em J. Opt. Soc. Am.}, vol.~67, no.~9, pp.~1252--1261, 1977.

\bibitem{Husimi:1940:PPMSJ:264}
K.~Husimi, ``Some formal properties of the density matrix,'' {\em Proc. Phys.
  Math. Soc. Jpn.}, vol.~22, no.~4, pp.~264--314, 1940.

\bibitem{Wigner:1932:PR:749}
E.~Wigner, ``On the quantum correction for the thermodynamic equilibrium,''
  {\em Phys. Rev.}, vol.~40, no.~5, pp.~749--759, 1932.

\bibitem{Cohen:1995}
L.~Cohen, {\em Time-Frequency Analysis}.
\newblock New Jersey: Prentice Hall PTR, 1995.

\bibitem{Auger:2008:131}
F.~Auger and E.~Chassande-Mottin, {\em Quadratic Time-Frequency Analysis I:
  Cohen's Class}, pp.~131--163.
\newblock 2008.

\bibitem{Wollenhaupt:2010:PSUPSNMXIV:76000X}
M.~Wollenhaupt, L.~Englert, A.~Horn, and T.~Baumert, ``Temporal femtosecond
  pulse tailoring for nanoscale laser processing of wide-bandgap materials,''
  {\em Proc.Spie, Ultrafast Phenomena in Semiconductors and Nanostructures
  Materials XIV}, vol.~7600, pp.~76000X--76000X--11, 2010.

\bibitem{Kerbstadt:2017:OE:12518}
S.~Kerbstadt, D.~Timmer, L.~Englert, T.~Bayer, and M.~Wollenhaupt, ``Ultrashort
  polarization-tailored bichromatic fields from a {C}{E}{P}-stable white light
  supercontinuum,'' {\em Opt. Express}, vol.~25, no.~11, p.~12518, 2017.

\bibitem{Kerbstadt:2017:JMO:1010}
S.~Kerbstadt, L.~Englert, T.~Bayer, and M.~Wollenhaupt, ``Ultrashort
  polarization-tailored bichromatic fields,'' {\em J. Mod. Opt.}, vol.~64,
  no.~10-11, p.~1010, 2017.

\bibitem{Baumert:1997:APB:779}
T.~Baumert, T.~Brixner, V.~Seyfried, M.~Strehle, and G.~Gerber, ``Femtosecond
  pulse shaping by an evolutionary algorithm with feedback,'' {\em Appl. Phys.
  B}, vol.~65, no.~6, pp.~779--782, 1997.

\bibitem{Yelin:1997:OL:1793}
D.~Yelin, D.~Meshulach, and Y.~Silberberg, ``Adaptive femtosecond pulse
  compression,'' {\em Opt. Lett.}, vol.~22, no.~23, pp.~1793--1795, 1997.

\bibitem{Koehnke:2023:NJP:123025}
D.~K\"{o}hnke, K.~Eickhoff, T.~Bayer, and M.~Wollenhaupt, ``Multichromatic
  supercontinuum polarization shaping applied to photoelectron holography,''
  {\em New J. Phys.}, vol.~25, no.~12, p.~123025, 2023.

\bibitem{Kaneyasu:2026:PRA:043104}
T.~Kaneyasu, Y.~Hikosaka, S.~Wada, M.~Fujimoto, H.~Iwayama, H.~Ota, K.~Shimizu,
  M.~Katoh, and F.~Koike, ``Observing the temporal evolution of a core hole
  with photoelectron interferometry,'' {\em Phys. Rev. A}, vol.~114, no.~4,
  p.~043104, 2026.

\bibitem{Prince:2016:NP:176}
K.~C. Prince, E.~Allaria, C.~Callegari, R.~Cucini, G.~De~Ninno, S.~Di~Mitri,
  B.~Diviacco, E.~Ferrari, P.~Finetti, D.~Gauthier, L.~Giannessi, N.~Mahne,
  G.~Penco, O.~Plekan, L.~Raimondi, P.~Rebernik, E.~Roussel, C.~Svetina,
  M.~Trov\`{o}, M.~Zangrando, M.~Negro, P.~Carpeggiani, M.~Reduzzi, G.~Sansone,
  A.~N. Grum-Grzhimailo, E.~V. Gryzlova, S.~I. Strakhova, K.~Bartschat,
  N.~Douguet, J.~Venzke, D.~Iablonskyi, Y.~Kumagai, T.~Takanashi, K.~Ueda,
  A.~Fischer, M.~Coreno, F.~Stienkemeier, Y.~Ovcharenko, T.~Mazza, and
  M.~Meyer, ``Coherent control with a short-wavelength free-electron laser,''
  {\em Nature Photonics}, vol.~10, no.~3, pp.~176--179, 2016.

\bibitem{Cattaneo:2016:OE:29060}
L.~Cattaneo, J.~Vos, M.~Lucchini, L.~Gallmann, C.~Cirelli, and U.~Keller,
  ``Comparison of attosecond streaking and rabbitt,'' {\em Opt. Express},
  vol.~24, no.~25, pp.~29060--29076, 2016.

\bibitem{Richter:2024:Nature:337}
F.~Richter, U.~Saalmann, E.~Allaria, M.~Wollenhaupt, B.~Ardini, A.~Brynes,
  C.~Callegari, G.~Cerullo, M.~Danailov, A.~Demidovich, K.~Dulitz, R.~Feifel,
  M.~D. Fraia, S.~D. Ganeshamandiram, L.~Giannessi, N.~G\"{o}lz, S.~Hartweg,
  B.~von Issendorff, T.~Laarmann, F.~Landmesser, Y.~Li, M.~Manfredda,
  C.~Manzoni, M.~Michelbach, A.~Morlok, M.~Mudrich, A.~Ngai, I.~Nikolov,
  N.~Pal, F.~Pannek, G.~Penco, O.~Plekan, K.~C. Prince, G.~Sansone,
  A.~Simoncig, F.~Stienkemeier, R.~J. Squibb, P.~Susnjar, M.~Trovo, D.~Uhl,
  B.~Wouterlood, M.~Zangrando, and L.~Bruder, ``Strong-field quantum control in
  the extreme ultraviolet domain using pulse shaping,'' {\em Nature},
  vol.~636, no.~8042, pp.~337--341, 2024.

\end{thebibliography}
\end{document}